\documentclass{article}

\usepackage{PRIMEarxiv}

\usepackage[utf8]{inputenc} 
\usepackage[T1]{fontenc}    
\usepackage{hyperref}       
\usepackage{url}            
\usepackage{booktabs}       
\usepackage{longtable}      
\usepackage{amsfonts}       
\usepackage{nicefrac}       
\usepackage{microtype}      
\usepackage{lipsum}
\usepackage{fancyhdr}       
\usepackage{graphicx}       
\usepackage{appendix}
\graphicspath{{media/}}     

\title{Embedded Off-Switches for AI Compute}

\author{
  James Petrie  \thanks{james.petrie@protonmail.com} \\
  \\
}

\begin{document}
\maketitle

\begin{abstract}
To address the risks of increasingly capable AI systems, we introduce a hardware-level off-switch that embeds thousands of independent ``security blocks'' in each AI accelerator. This massively redundant architecture is designed to prevent unauthorized chip use, even against sophisticated physical attacks. Our main security block design uses public key cryptography to check the authenticity of authorization licenses, and randomly generated nonces to prevent replay attacks. We evaluate attack vectors and present additional security block variants that could be added for greater robustness. Security blocks can be built with standard circuit components, ensuring compatibility with existing semiconductor manufacturing processes. With embedded security blocks, the next generation of AI accelerators could be more robustly defended against dangerous misuse.
\end{abstract}


\section{Introduction}

AI systems are becoming rapidly more capable, which brings unprecedented dangers from misuse, loss of control \cite{bengio_international_2025}, and gradual disempowerment \cite{kulveit_gradual_2025}. Compute is essential for AI development, so governance over compute is a promising way to govern AI development \cite{sastry_computing_2024}. While direct physical control of data centers is the most straightforward governance method, economic pressures and strategic interests often lead to the deployment of high-performance chips in locations with limited oversight, creating vulnerabilities like theft and unauthorized use.

An off-switch built directly into chips could make governance much more robust against chip theft and chip diversion \cite{aarne_secure_2024, kulp_hardware-enabled_2024, ramiah_toward_2025}. This would be especially valuable if the AI risk landscape changes rapidly \cite{barnett_ai_2025} (e.g. if a frontier lab finds that a typical AI development strategy leads to power-seeking AI systems that are capable of developing bioweapons). Hardware-based approaches could also enable distributed governance models requiring cryptographic consensus from multiple parties before authorizing chip usage \cite{petrie_flexible_2025}.

In previous work, we proposed a firmware-based off-switch that uses existing chip security hardware as the fastest way to deploy a hardware-backed off-switch \cite{petrie_near-term_2024}. While firmware-based mechanisms provide near-term solutions, they could be vulnerable to sophisticated physical attacks targeting single points of failure. 

In this paper, we propose a significantly more secure design that would embed thousands of hardware security blocks into upcoming generations of AI chips. Each security block would individually check for up-to-date usage authorization, and block essential chip operations if this approval is not received. Cyber defense is typically difficult because a defender needs to win every single time, while an attacker only needs to win once. For the particular problem of chip locking, this heuristic can be reversed because an attacker would need to circumvent every single security block without damaging the chip.

Modern chip design uses many third-party IP blocks, so it would be a standard procedure to add security blocks. Each security block could be built with less than 40,000 transistors, so even 10,000 security blocks would take up a very small area relative to the rest of an AI chip (e.g. $<1\%$ for a Nvidia H100 with 80 billion transistors).

We begin by presenting the high level security block architecture in Section \ref{sec:design}, then the primary design based on random number generation and public key cryptography in Section \ref{sec:primary_design}. In Section \ref{sec:threat} we analyze potential attacks, and in Section \ref{sec:alternative_designs}, we discuss alternative security block designs that could be added in parallel. 

\section{High-Level Design}
\label{sec:design}

\begin{figure}[h]
\begin{center}
  \includegraphics[width=12cm]{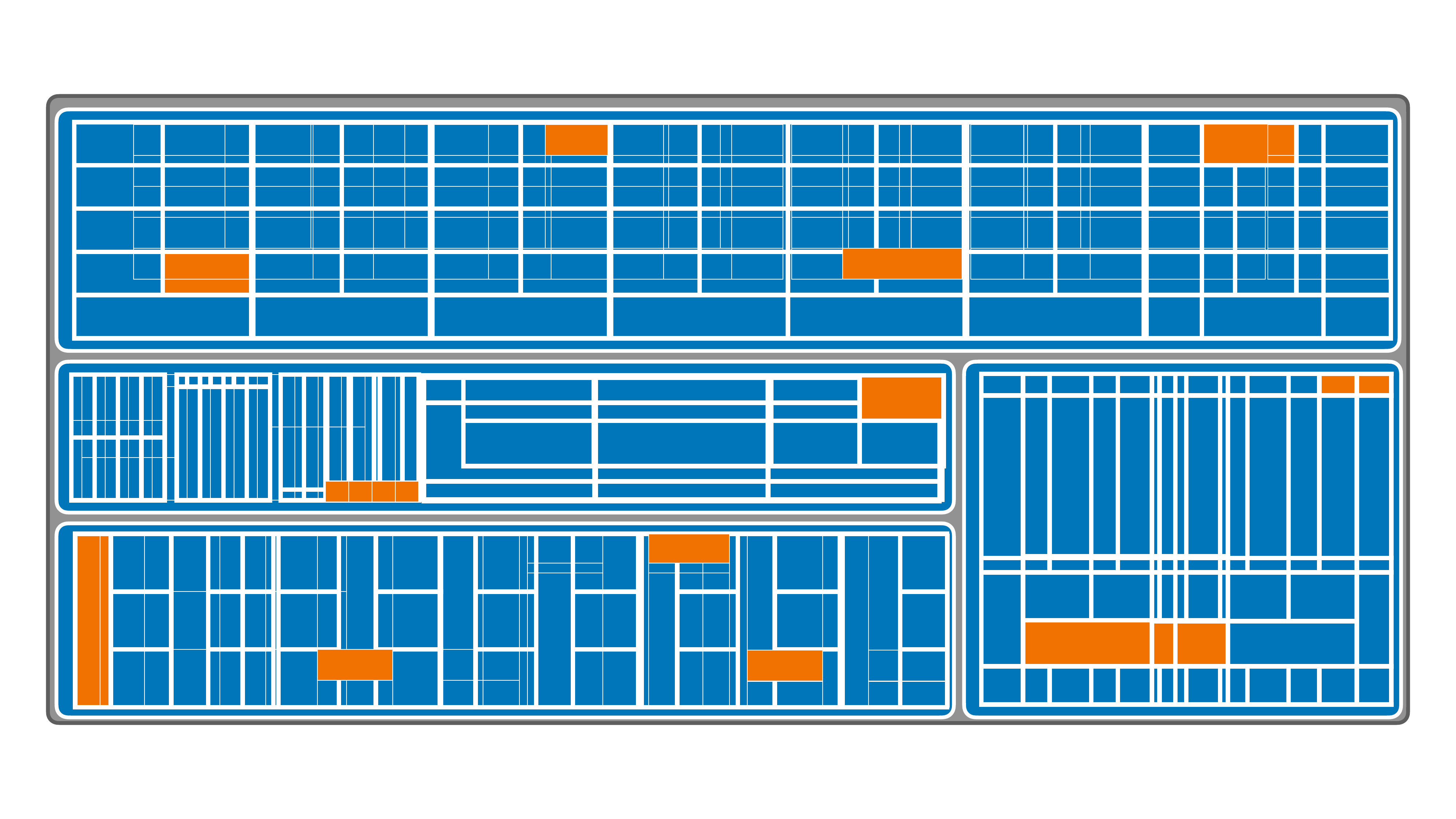}
  \caption{Conceptual diagram of an AI accelerator with thousands of security IP blocks (orange) interspersed with computational logic (blue), ensuring all critical data paths require authorization.}
  \label{fig:chip_architecture}
\end{center}
\end{figure}

The proposed design integrates thousands of small security block circuits throughout each chip. Each security block acts independently and implements a `deadman's switch', where the security block will only enable essential logic if it has recently received authorization. A deadman's switch architecture is necessary because there is not a reliable way to send an `off' signal to chips in unknown locations that are physically shielded and not connected to the internet.

\begin{figure}[h]
\begin{center}
  \includegraphics[width=12cm]{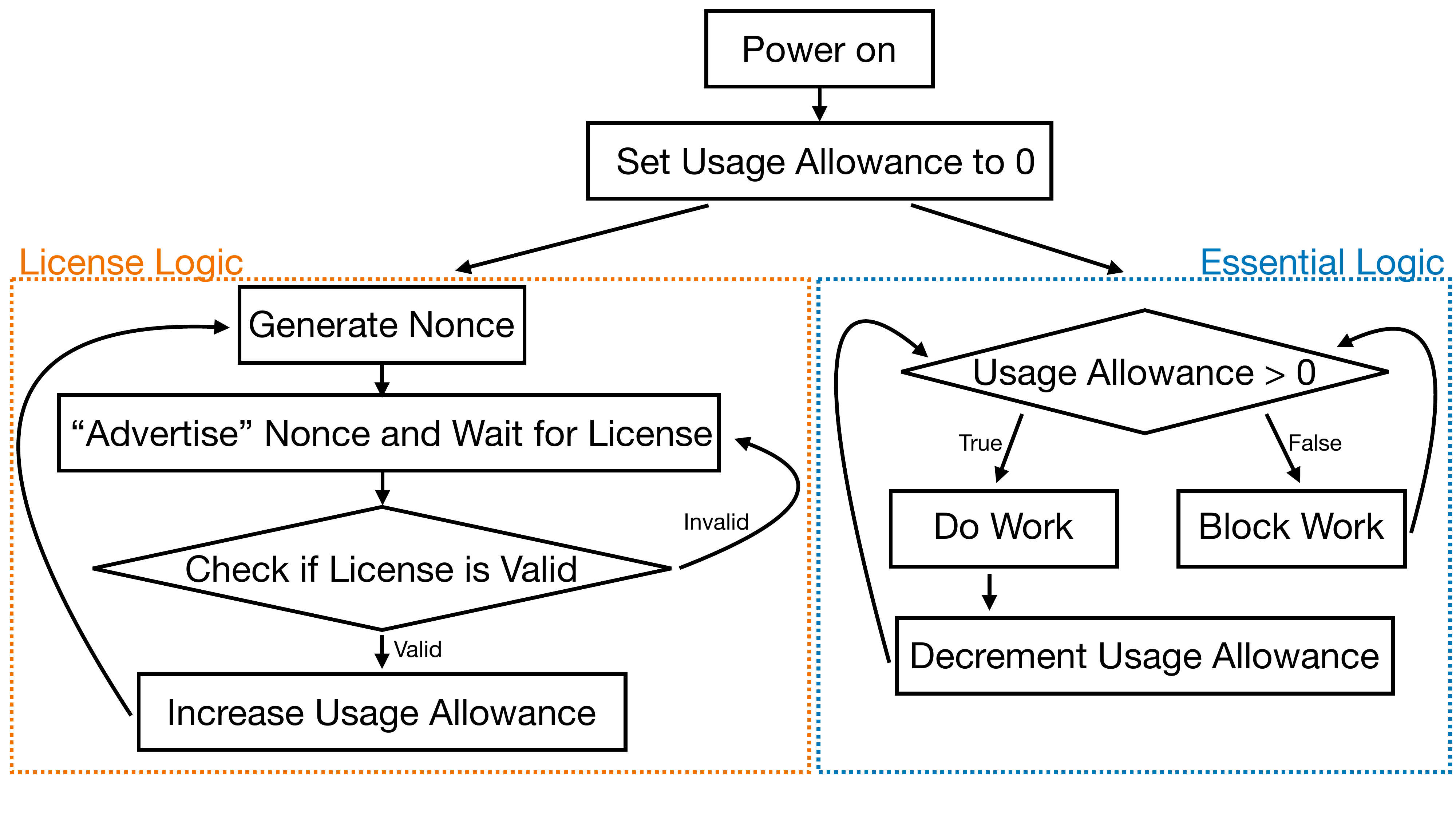}
  \caption{High-level control flow for a security block that prevents essential logic form functioning unless a valid license has been processed recently.  }
  \label{fig:control_flow}
\end{center}
\end{figure}

The process for receiving updated usage approval begins when a security block generates a unique `nonce' and sends it to an authorizer (likely via the chip's owner). To approve usage, the authorizer uses a key to generate a `license' valid for that specific nonce and returns it. Finally, the security block cryptographically verifies the license's authenticity and, if valid, increments its stored `usage allowance' value. A flow chart illustrating this process is shown in Figure \ref{fig:control_flow}.

The usage allowance stored in each security block has the following properties:
\begin{enumerate}
    \item Initializes to 0 on chip power-on.
    \item Increments by a specified amount upon receipt of a valid license.
    \item Decrements by one for each authorized computational operation.
    \item Halts the essential logic it controls if the allowance reaches 0.
\end{enumerate}

Each of the security blocks would be made responsible for a simple but critical part of the chip's regular operations, and refuse to perform that operation if it does not have ongoing authorization. For example, some blocks could be responsible for addition logic or as switches for message passing. If any local parts of an AI chip are broken, the overall computation would be faulty. So, to use a chip, an attacker would need to disable the majority of security blocks without damaging the standard logic they are typically responsible for.

Chips would also need logic to collect nonces from security blocks and export these, and also to return licenses from authorizers to the appropriate security blocks. The security requirements for this communication logic are less demanding, because everything external to the security block is considered to be untrusted. This collection and communication process could potentially be managed by an on-chip microcontroller like the RISC-V cores on Nvidia chips.

\section{Main Security Block Design: Public-Key Cryptography and TRNG }
\label{sec:primary_design}

In this section, we present our main design for the security block that is based on randomized nonce generation and public key license validation. In Section \ref{sec:alternative_designs}, we sketch other potential designs that make different security assumptions, including options that use symmetric key cryptography, or no cryptography at all. Multiple types of designs can be built into each chip for more robust defense against attackers with unknown capabilities.

\subsection{Randomized Nonce Generation}

Each block can generate a unique nonce using an on-chip True Random Number Generator (TRNG), implemented with a ring oscillator requiring only a small number of logic gates (e.g. 70 transistors for Vasyltsov et al's design \cite{vasyltsov_fast_2008}). With a 128-bit nonce, the probability of matching a prior approved license is statistically negligible ($\sim 10^{-25}$, assuming $36 \times 10^{12}$ prior licenses\footnote{Number of prior approved licenses: $10^3$ blocks/chip $\times 10^6$ chips $\times 2$ licenses/(block $\times$ day) $\times 5$ years $\times 365$ days/year $\sim 36 \times 10^{12}$. $36 \times 10^{12}/2^{128} = 10^{-25}$.}). Entropy can be XORed from multiple sources to defend against attacks that attempt to bias the nonce generation.

\subsection{License Validation with Public Key Cryptography}

Public key cryptography allows the holder of a private key to sign a message. Holders of the associated public key can check that the signature is authentic. For our design, the public key(s) for license authentication are hardwired into the circuit (Mask ROM). This hardwiring reduces the risk of accidental or intentional misconfiguration that might be possible if the public key(s) were instead written to antifuses after manufacturing. The corresponding private key(s) would be stored securely off-device by the authorizer. By using public key cryptography, this design is not susceptible to scanning or sidechannel attacks (because no secrets would be stored on-device). 

Elliptic Curve Digital Signature Algorithm (ECDSA) is an extensively used and validated public key algorithm. Compact ECDSA circuits have been developed with less than 9000 gates \cite{wenger_hardware_2011}. 

\subsection{Essential Logic and Physical Integration}

There is a lot of flexibility about what essential logic should be performed by the security block, so long as it is critical for proper chip functioning. This allows placement of security blocks in areas that are less sensitive to chip design constraints. To avoid timing closure constraints \cite{chadha_static_2009}, security blocks could be placed in parts of pipelines with more timing slack. As a minimum example, the essential logic could be a "switch" circuit that is responsible for routing data packets in different directions (e.g. if the first bit in the address is 0 it sends data left and if 1 sends data right). The transistors for essential logic within the security block can be placed directly next to critical security logic to make tampering more risky.

The number of gates is mainly determined by the public key circuitry because the essential and security logic is relatively simple, and a ring oscillator can be implemented with < 100 standard gates \cite{vasyltsov_fast_2008}. A compact ECDSA implementation requires roughly 9000 gates. Using a standard 4-transistor-per-gate conversion, this translates to 40,000 transistors per block. For context, this is a negligible fraction ($5 \times 10^{-7}$) of the 80 billion transistors in an NVIDIA H100 GPU. Even with 10,000 security blocks, the total area overhead would be approximately $0.5\%$ of the die.

\begin{figure}[h]
\begin{center}
  \includegraphics[width=12cm]{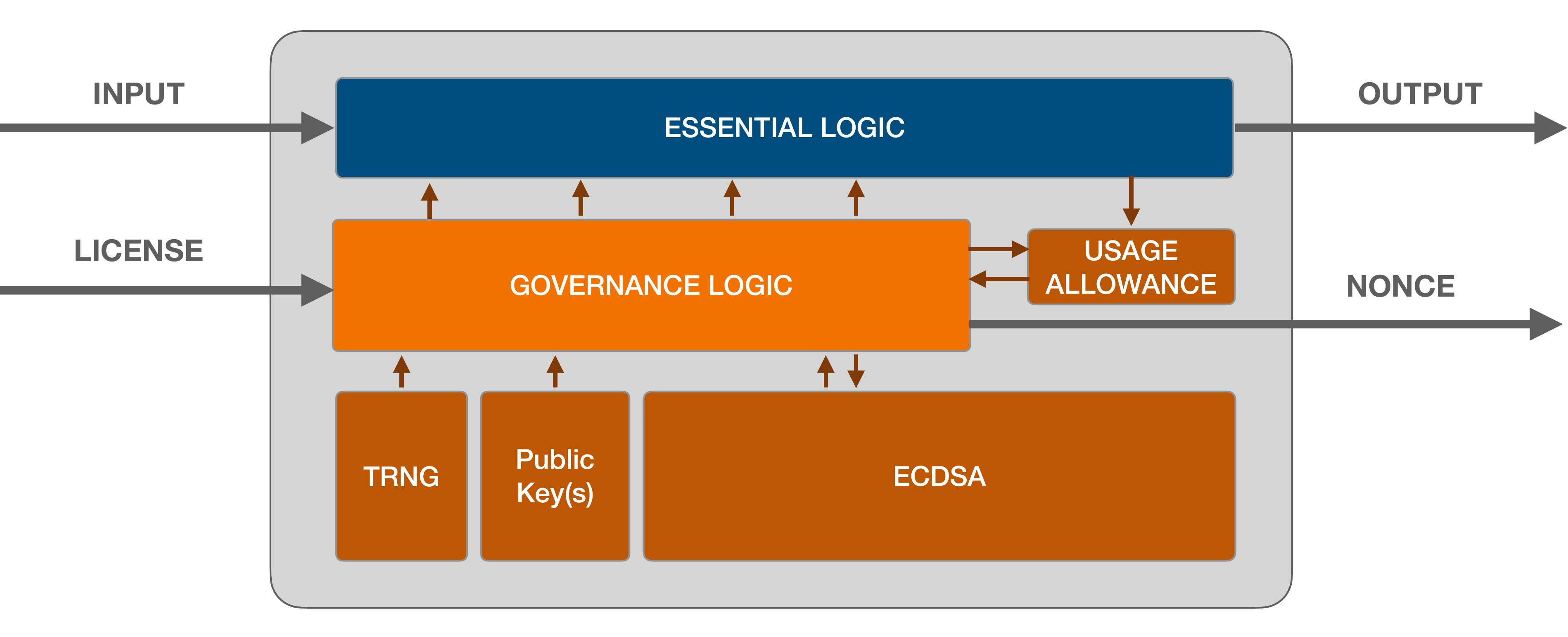}
  \caption{Conceptual diagram of security block. The actual layout could be modified to put the most security-critical transistors close to essential logic. ECDSA: Elliptic Curve Digital Signature Algorithm, TRNG: True Random Number Generator. }
  \label{fig:security_block}
\end{center}
\end{figure}

\section{Threat Model}
\label{sec:threat}

\subsection{Security Goals}

The primary security objective is to protect against unauthorized usage\footnote{How authorization decisions are made is out of scope for this paper, but some options for authorizing parties are domestic regulators or countries that are part of an international agreement.} of large numbers of chips (10,000+). The system must also prevent the security blocks themselves from being exploited to block legitimate usage of chips, or secretly exfiltrate data from chips or manipulate computations.

\subsection{Threat Actors}

The primary threat actor is a well-funded, non-state entity (e.g., a clandestine \$1B project) with significant hardware expertise aiming to covertly unlock chips at scale.

The ambitious threat model is a nation-state actor with physical access to chips overtly attempting to unlock them. Nation-state actors may be involved in chip design and manufacturing, may have secret information about cryptographic exploits (e.g., weaknesses in RSA or other algorithms), and may have a large amount of existing expertise and tooling to attempt sophisticated physical exploits.

\subsection{Potential Attacks and Countermeasures}

\begin{longtable}{p{3cm}|p{5.5cm}|p{6.5cm}}
\caption{Hardware Attack Vectors and Countermeasures} \label{table:attacks} \\
\toprule
\textbf{Attack Vector} & \textbf{Description} & \textbf{Countermeasures} \\
\midrule
\endfirsthead
\multicolumn{3}{c}%
{{\bfseries \tablename\ \thetable{} -- continued from previous page}} \\
\toprule
\textbf{Attack Vector} & \textbf{Description} & \textbf{Countermeasures} \\
\midrule
\endhead
\midrule \multicolumn{3}{r}{{Continued on next page}} \\
\endfoot
\bottomrule
\endlastfoot

Logical Flaws & Design flaws in a security block could allow circumvention without physical modification, such as a timing glitch that incorrectly modifies the usage allowance. & 1. Use multiple, independently designed security blocks on each chip, requiring an attacker to find unique exploits for each type. \par 2. Prioritize minimal, auditable designs and use formal verification to eliminate logical bugs.  \\
\midrule
License Reuse & If an attacker could reuse a valid license multiple times this would grant unlimited chip usage. & Validating that the license is correct for the current nonce, and only using unique nonces defends against this attack. Two ways to ensure nonce uniqueness are: \par 1. Persistent Memory: generate a unique nonce by combining a unique block ID and a license counter that are stored in nonvolatile within-block memory (e.g., antifuses). Increment the license counter each time a license is used\footnote{Although writing to antifuses depends on an external high-voltage source that may be vulnerable to disablement.}. \par 2. True Random Number Generator (TRNG): Use an on-block TRNG to generate a large random number (e.g., 128-bit), making nonce repetition highly unlikely. Ring oscillator TRNGs only require a small number of standard gates \cite{vasyltsov_fast_2008}. Multiple entropy sources can be XOR'd together for better reliability. \\
\midrule
Execution Bypass & An attacker modifies their workload to use only computational pathways on the chip that are not gated by any security blocks. & Audit the physical chip layout to ensure all computational and data routing paths are gated by several security blocks. \\
\midrule
Voltage or Laser Glitching & Voltage or laser glitching attempts to skip critical security checks by inducing logic errors. Such attacks could target the license verification process in security blocks. & 1. Glitches against a single security mechanism are usually unreliable. With a large number of security blocks, an attacker would need to successfully glitch thousands of physically separate circuits. \par 2. Randomizing the timing for critical checks across blocks would prevent synchronized attacks. \par 3. Glitch detector circuits \cite{gomina_power_2014} could be integrated into security blocks to reset the usage counter if this type of attack is detected.  \\
\midrule
Physical Tampering & Tools like a Focused Ion Beam (FIB) could physically edit circuits, for example, to bypass a security block or hardcode a valid state. & 1. Advanced nodes (e.g., $<7$nm) that AI accelerators are fabricated on have a feature size that is smaller than the resolution of most common FIB editing tools (e.g. Gallium FIB \cite{scheffler_patterning_2016}). \par  2. With thousands of distributed security blocks, thousands of high-precision edits would be needed per chip, making a large-scale attack extremely complex.\par 3. Place security logic transistors immediately adjacent to the essential logic they control, making any edit attempt risk damaging the chip's functionality. \par 4. Where possible, design security blocks with internal redundancy so that a single circuit edit would not be sufficient (e.g. by having multiple redundant counters for usage allowance). \par 5. Incorporate security blocks into physically shielded or hard-to-access components, such as logic layers within an HBM stack. \par 6. Traditional physical security measures like secure enclosures \cite{anderson_security_nodate}. \\
\midrule
Secret Extraction & An attacker uses physical probing, side-channel analysis (e.g., power monitoring, EM emissions), or other methods to extract on-chip cryptographic secrets. & 1. With public-key cryptography, only the public key needs to be stored on the chip, so there is no secret to extract. \par 2. If secrets are stored on-chip, make them unique so that scanning one block doesn't unlock many. \par 3. Substantially increase the number of secrets that an attacker would need to extract by using a large number of security blocks. \par 4. Defend against side channels by using constant-time algorithms and other hardware-level countermeasures that normalize power consumption and EM signatures. \\
\midrule
Supply Chain Compromise & A malicious actor introduces a hardware backdoor during the chip design or manufacturing phase to circumvent or misuse the security blocks. & Perform audits of the final design and physically deconstruct and image a statistical sample of manufactured chips to detect unauthorized modifications. \\
\midrule
Cryptographic Vulnerabilities & Vulnerabilities in cryptographic algorithms could make it possible for an attacker to generate valid licenses (e.g. by brute forcing the public key, potentially using quantum computers). & 1. Choose well-validated algorithms.\par 2. Use multiple security block designs with different cryptographic algorithms. \par 3. Some security blocks should be quantum resistant, and ideally some will not rely on cryptography at all. \\
\end{longtable}

In addition to attacks directly against the security blocks, the overall governance framework relies on the secure management of authorization keys and the communication channels for licenses. Table \ref{table:external_attacks} analyses attacks and countermeasures involving this external infrastructure. 

\begin{longtable}{p{3cm}|p{5.5cm}|p{6.5cm}}
\caption{External System Attack Vectors and Countermeasures} \label{table:external_attacks} \\
\toprule
\textbf{Attack Vector} & \textbf{Description} & \textbf{Countermeasures} \\
\midrule
\endfirsthead
\multicolumn{3}{c}%
{{\bfseries \tablename\ \thetable{} -- continued from previous page}} \\
\toprule
\textbf{Attack Vector} & \textbf{Description} & \textbf{Countermeasures} \\
\midrule
\endhead
\midrule \multicolumn{3}{r}{{Continued on next page}} \\
\endfoot
\bottomrule
\endlastfoot

Authorization Key Deletion & If an attacker could permanently destroy all keys used to generate licenses, chips would be permanently disabled. & 1. Store multiple, encrypted, and geographically distributed backups of the keys. \par 2. Use certified Hardware Security Modules (HSMs) for key storage and operations. \par 3. Implement robust physical access controls for all locations where keys are stored. \\
\midrule
Authorization Key Theft & An attacker gains access to the private keys, allowing them to generate unlimited valid licenses and completely bypass the off-switch mechanism. & 1. A quorum-based system (e.g., multi-signature) where multiple independent parties must cooperate to use the keys, prevents a single point of failure. \par 2. As with key deletion, use HSMs and strong physical access controls to prevent unauthorized access. \par 3. Use different authorization keys for different batches of chips to limit the scope of a single key compromise. \\
\midrule
Network Disruption & The communication channel between the chips (requesting licenses) and the authorization service is severed, preventing chips from receiving new licenses. & 1. Issue licenses with a sufficiently long validity period (e.g., days) to allow time to restore communication via alternate channels. \par 2. Plan for emergency out-of-band methods to transmit nonces and licenses, such as satellite links or even physical transport of storage devices. \\
\end{longtable}

\section{Additional Security Block Designs}
\label{sec:alternative_designs}

Each chip can incorporate several different security block designs, so no single vulnerability would compromise the whole chip. It may be worthwhile to generate multiple different physical layouts from the main RTL design. We can also keep the overall architecture, but replace the ECDSA cryptography with a variety of alternative public-key algorithms, ideally several of them quantum resistant.

By creating significantly different designs with different security assumptions, security is improved against sophisticated actors that may have unexpected capabilities (e.g. cryptography exploits). In this section, we briefly analyze additional designs that do not rely on random number generation, public key cryptography, or any cryptography at all.

\subsection{No TRNG}

The integrity of random number generators can be difficult to verify, so it would be useful to also have security blocks that don't rely on them. With persistent memory, a unique block ID can be appended to a license counter to create a unique nonce. Each time a license is used an antifuse bit must be written, so for daily licenses over 5 years, 2000 antifuse bits would be sufficient. Antifuse memory can be programmed within the security block by using an externally generated high voltage source (probably with a charge pump that is shared between multiple security blocks).

\subsection{No Public Key Cryptography}

Commonly used public key cryptography algorithms may be vulnerable to brute forcing by quantum computers, and quantum resistant public key algorithms are still undergoing review. These risks would be reduced by adding some symmetric key security blocks.

A symmetric approach pre-shares 256-bit keys stored in antifuses, using TRNG-generated nonces with AES-256 encryption for signatures. A side-channel resistant AES core occupies $20 \mu m^2$ on a 40nm process node \cite{zhang_compact_2018}. Scaling to a modern 4nm node, this area would be negligible, allowing tens of thousands of such blocks to be integrated while consuming less than $1\%$ of the total die area of a modern accelerator.

\subsection{No Cryptography}

To defend against vulnerabilities in cryptographic circuits and/or cryptographic algorithms, it is possible to build a security block that only depends on a large pre-shared secret. Physical unclonable functions (PUFs) \cite{herder_physical_2014} could be used for this type of security block, but for simplicity we will focus on an antifuse implementation here.

Each security block is programmed with $N$ antifuse bits preshared with the authorizer\footnote{The antifuse memory can be programmed while still in a trusted facility, so is not vulnerable to the previously discussed attacks against the voltage source}. Randomly generated nonces specify which $k$ bits a valid license must contain. To reduce the number of antifuses needed, a monthly licensing period can be used for these blocks (instead of daily) would require roughly 100 licenses over 5 years (assuming the chips are not frequently reset). With 50 bits revealed per license, these provided licenses would reveal at worst 5000 bits. Using N=10,000 total bits of secret data, then after 5 years half of the secret would be revealed and an attacker would need to make roughly $2^{50/2} = 2^{25}$ guesses to spoof a license. With 1-second delays added to nonce generation and license validation, this would provide approximately 1 year of brute-force resistance\footnote{Calculation: $2^{25}$ seconds $\approx 1$ year.}.

\section{Discussion}
\label{sec:discussion}

Embedded off-switches offer a promising path toward hardware-enforced AI governance. By deeply integrating thousands of redundant and diverse security blocks throughout AI accelerators, this architecture provides strong resistance to sophisticated circumvention attempts. The proposed approach is compatible with existing semiconductor design processes and could significantly strengthen oversight of frontier AI development.

Significant engineering and policy challenges remain. Hardware typically takes years to design and manufacture, and longer still for new AI accelerators to displace older ones \cite{emberson_stock_2025}. Integration into accelerators requires nontrivial support and work from leading accelerator designers. Communication of nonces requires $\sim 10$kB data output, which may raise concerns about data exfiltration, especially for airgapped datacenters. Third party trust in the security relies on detailed investigation into the design, manufacturing, and key provisioning. 

While the addition of thousands of security blocks uses minimal die area, it could complicate the physical design phase. Flaws in security blocks could block legitimate chip usage, though this risk can be mitigated with the same techniques that are successfully used for the many other IP blocks in a chip.

Beyond a simple on/off mechanism, this distributed security architecture could provide a foundation for more sophisticated hardware-enforced AI governance. Future work could extend this framework to enable fine-grained workload authorization, where licenses permit computation only for specific model weights or approved workloads. The blocks could also generate a tamper-resistant, distributed audit log, with each block cryptographically attesting to the computations it has gated. This would create a verifiable record of chip activity. Furthermore, the authorization process could be expanded to support multi-party governance schemes, requiring cryptographic consensus from multiple independent entities before chip usage is approved. 



\bibliographystyle{unsrt}  
\bibliography{OffSwitch}

\end{document}